\documentclass[lettersize,journal,twocolumn]{IEEEtran}
\usepackage{amssymb,amsmath}
\usepackage{cite}
\usepackage{graphicx}
\usepackage{psfrag}
\usepackage{url}
\usepackage[T1]{fontenc}
\usepackage[utf8]{inputenc}
\usepackage[absolute,overlay]{textpos}
\usepackage{gensymb}
\usepackage{cases}
\usepackage{cuted}
\usepackage[font=footnotesize]{caption}
\usepackage[font=footnotesize]{subcaption}
\usepackage{float}
\usepackage[linesnumbered,ruled,lined]{algorithm2e}
\SetKwInOut{Training}{Train}
\SetKwInOut{Testing}{Test}
\usepackage{hyperref}
\usepackage{pgf}

\usepackage[nocomma]{optidef}
 \newcommand\modification[1]{\textcolor{black}{#1}}

\usepackage{amsthm}

\newtheorem{remark}{Remark}

\usepackage{booktabs}
\usepackage{color,soul}

\usepackage{textcomp}
\usepackage{xcolor}
\def\BibTeX{{\rm B\kern-.05em{\sc i\kern-.025em b}\kern-.08em
    T\kern-.1667em\lower.7ex\hbox{E}\kern-.125emX}}

\usepackage{tabularx}
\usepackage{makecell}
\usepackage{multirow}

\usepackage{graphicx}
\usepackage{caption}
\usepackage{subcaption}

\begin{document}

\title{Wireless Energy Transfer from Space to Ground via Satellite Constellation Grids}
\author{
	\IEEEauthorblockN{Mohammad Shehab,~\IEEEmembership{Member,~IEEE}, Osmel M. Rosabal,~\IEEEmembership{Member,~IEEE}, Onel L. A. L\'opez,~\IEEEmembership{Senior Member,~IEEE} and Mohamed-Slim Alouini,~\IEEEmembership{Fellow,~IEEE} \\
	}
	\thanks{This work is partially supported by the KAUST Office of Sponsored Research under Award ORA-CRG2021-4695 and the Academy of Finland, 6G Flagship program (Grant no. 346208)}
	
	\thanks{Mohammad Shehab and Mohamed-Slim Alouini are with CEMSE Division, King Abdullah University of Science and Technology
(KAUST), Thuwal 23955-6900, Saudi Arabia. Emails: mohammad.shehab@kaust.edu.sa, slim.alouini@kaust.edu.sa

Osmel M. Rosabal and Onel L. A. L\'opez are with Centre for Wireless Communications (CWC), University of Oulu, Finland. Email: firstname.lastname@oulu.fi. 

}
}
\maketitle

\begin{abstract}

This letter presents a framework for space-to-ground wireless energy transfer (WET) for wirelessly chargeable devices (WCD) located in remote areas or disaster situations. We consider a grid of multi-antenna satellites that charge a WCD within line-of-sight. Closed-form expressions for harvested energy are derived considering maximum ratio transmission (MRT) ensuring that the WCD meets its circuit charging threshold $P_{th}$. Simulations elucidate that milli-joule-level energy can be harvested during satellite grid visibility, with charging efficiency influenced by the number of satellites, their altitude, charging frequency, and grid inclination. 

All codes used in this paper can be found from:
\url{https://github.com/moshehab570/Satellite_WET.git}


\end{abstract}

\section{Introduction}\label{sec:introduction}

In disaster-affected or remote regions such as vast oceans and deserts, access to stable power supplies is often disrupted or entirely unavailable, severely limiting the operation of critical communication systems, sensors, and rescue equipment. Traditional power delivery methods—such as wired connections or manual battery replacement can be slow, dangerous, or logistically impossible to deploy in such conditions. \modification{WET} emerges as a fast and reliable solution, enabling the remote powering or recharging of devices without physical contact or human presence.

Using WET, low-power wireless and electronic devices can be recharged remotely using terrestrial power beacons, aerial platforms, or satellites. For instance, in disaster situations such as earthquakes and floods, a wirlessly chargeable phone with a depleted battery could be wirelessly powered, allowing a victim to make emergency calls, share injury details, or help rescuers locate them under debris. Additionally, WET can reactivate sensors and cameras in power-limited areas to retrieve crucial data about past events \cite{disaster}.

\modification{To this end, terrestrial or aerial WET providers might be infeasible, costly, or slow to deploy in remote and disaster-stricken areas. Nowadays, the availability of thousands of low Earth orbit (LEO) satellites plays an important role in providing connectivity to off-grid isolated regions and to civilians affected by disasters such as earthquakes, floods, and military conflicts. These scenarios can exploit the wide satellite coverage and near-global visibility to deliver milli-watt-level power, which is sufficient to operate low-power devices and to support user devices for short periods during emergencies. However, research on delivering energy from satellite constellations to devices in such conditions remains limited.}

In this context, the work in \cite{Zarini} discussed an age-optimized simultaneous wireless information
and power transfer for energy-limited devices. Their work provided useful insights on the age-energy tradeoff from queuing theory perspective, while providing solution based on meta-learning. However, it did not include closed-form solutions for the harvested energy or how the system would scale up with multiple satellites and antennas. The authors of \cite{Wang.2022} discussed stochastic geometry approaches for the analysis of dense LEO satellite systems. Meanwhile, \cite{Su.2019} provided an overview of architecture and key technologies, including beam design for satellite constellations and the impact on system performance indicators such as coverage. Furthermore, \cite{Heo.2023} presented a detailed survey of MIMO satellite systems.

The authors of \cite{Xu.2023} proposed a novel distributed beamforming technology to enhance the space-to-ground link budget based on the superposition of electromagnetic waves radiated from multiple satellites. Meanwhile, the works in \cite{Kyriatzis.2025,Qaraqe.2022} discussed inter-satellite optical power transfer, and the authors of \cite{Zhu.2024} illustrated a movable antenna setup to significantly reduce interference leakage.


\modification{To this end, we promote the idea of providing energy from space to low-power IoT or end-user devices located in remote and disaster areas, where access to power supplies might be challenging. We start by deriving the harvestable energy at a WCD within the charging visibility range of a grid of satellites equipped with multiple antennas. The obtained expresssions account for the azimuth angle of inclination of the satellite orbit plane with respect to the WCD, the WCD circuit threshold, as well as the MRT gain of the multi-antennas at the satellite grid. We present extensive numerical results to affirm that it is possible to harvest a reasonable amount of energy during the visibility period of a satellite grid, specially for higher number of satellites and antennas. The results highlight the harvestable energy characterization for different satellite heights, carrier frequencies, WCD circuit threshold values, and satellites' azimuth inclination angles. Interestingly, a WCD can harvest more than 10 mJ from a grid of only 10 satellites at the low frequency microwave range (i.e, few hundred MHz), thanks to the MRT gain.}


\section{System Layout and Analysis}\label{sec:sys}
\subsection{System Model}
\begin{figure}[t!]
\centering
\includegraphics[width=0.98\linewidth]{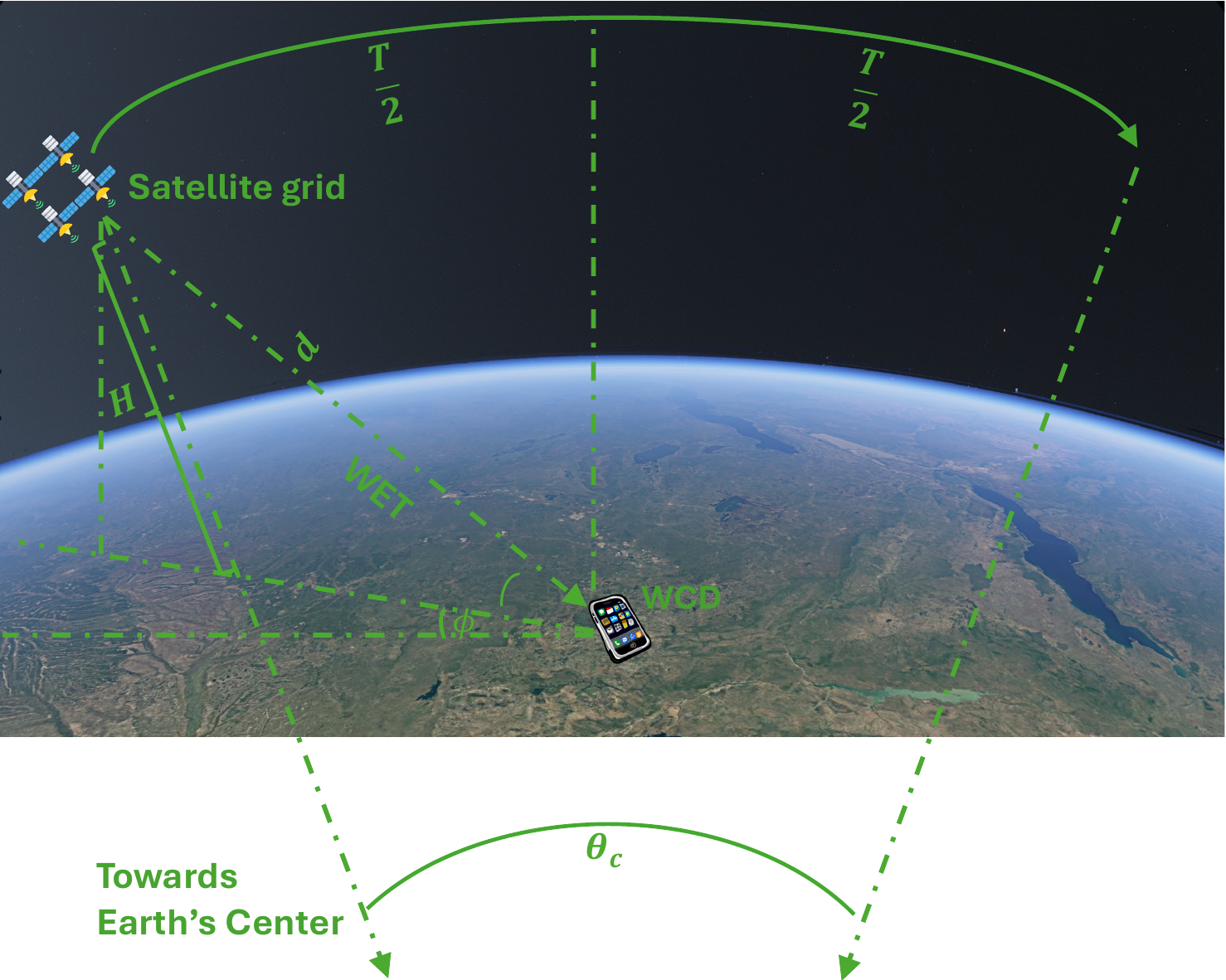} \vspace{1mm}
\caption{A satellite grid transferring energy to a WCD.}
\label{fig:sys}
\end{figure}

Consider a WCD with a charging duration $T$, which spans the time of rotation of the closest satellite grid from the start until the end of the charging visibility range of the WCD during the energy harvesting process. During this duration, the device can observe the satellite grid, which constitutes $N$ constellation satellites moving together as shown in Fig. \ref{fig:sys}. 

Assuming visibility and \modification{line of sight (LoS)}, the 
\modification{received power at the}
WCD from one satellite at time $t$ is given by Friss equation as \cite{Anna,Eslam1}
\begin{align} \label{p_r1}
P_r(t) = \mathbb{E}_{|h(t)|^2} \left[ P_t G_T G_R \frac{\lambda^2}{(4\pi d(t))^2} |h(t)|^2 \right],
\end{align}
where $\lambda=\frac{c}{f}$ is the charging carrrier wavelength at frequency $f$, $P_t$ is the transmit power, $G_T$, and $G_R$ are the transmit and receive antenna gains, respectively. $c$ is the speed of light, \modification{$|h(t)|^2$ is the satellite channel coefficients, which is a function of time and follows a shadowed Rician fading distribution}. According to \cite{distance}, the distance $d(t)$ between the satellite and the device is expressed as\footnote{The proof of this expression follows a straight forward manipulation of the law of cosines and is omitted due to page limit.}
\begin{align} \label{distance}
d(t)= \sqrt{(R + H)^2 + R^2 - 2 (R+H) R \cos \phi \cos(\omega t)},
\end{align}
where $R=6378$ km is the earth's radius, and $H$ is the height of the satellite, and $\omega=\sqrt{\frac{GM_e}{(R + H)^3}}$ (rad/s) is the satellite angular velocity with $G$ being the gravitational constant, and $M_e$ being the mass of Earth. According to \cite{Anna}, the shadowed Rician fading can be approximated by a gamma random variable, whose probability density function (pdf) is given by
\begin{align}
f_{|h|^2}(x)=\frac{1}{\Gamma(\alpha_s) \beta_s^{\alpha_s}} x^{\alpha_s-1} e^{-x/\beta_s},
\end{align}
where  
\begin{align}
\alpha_s = \frac{m(2b_0 + \Omega)^2}{4mb_0^2 + 4mb_0\Omega + \Omega^2},
\end{align}
\begin{align}
\beta_s = \frac{4mb_0^2 + 4mb_0\Omega + \Omega^2}{m(2b_0 + \Omega)}, 
\end{align}
where $m$, $b_o$, and $\Omega$ being the parameters of the shadowed Rician fading as follows: $m$ is the shadowing parameter, $b_0$ is the scatter component, and $\Omega$ is the LoS \modification{component}. Herein, $\Gamma(.)$ is the Gamma function. According to \cite{book}, the expectation of this distribution is $\alpha_s \beta_s$. Hence, the expression in \eqref{p_r1} can be written as
\begin{align} \label{p_r2}
P_r(t) =   P_t G_T G_R \frac{\lambda^2}{(4\pi d(t))^2}  \alpha_s \beta_s.
\end{align}

\subsection{Harvested Energy}
\modification{The harvested power at the WCD is modeled as
\begin{equation}\label{eq:ehModel}
    P_h(t) = \begin{dcases}
        0, &P_r(t) < P_{th}, \\
        \eta_hP_r(t), &P_r(t)\geq P_{th},
        \end{dcases}
\end{equation}
where $\eta_h$ is the conversion efficiency of the circuit. Herein, we assume that the WCD operates predominantly close to its sensitivity threshold $P_{th}$ as expected in satellite-enabled WET scenarios. Therefore, the active harvesting region is approximated by a linear function.}

We define the cut-off angle $\theta_c$ as the maximum angle at which the received power at the WCD is above the circuit sensitivity (i.e, $P_r\geq P_{th}$). From \eqref{distance}, $\theta_c$ is given by
\begin{align} \label{theta_th}
\theta_c=  \cos^{-1} \left( \frac{(R + H)^2 + R^2 - d_{c}^2}{2 (R + H) R \cos\phi} \right),
\end{align}
with
\begin{align} \label{d_th}
d_{c} = \frac{\lambda}{4\pi} \sqrt{ \frac{P_t G_T G_R \, \alpha_s \beta_s}{P_{th}} },
\end{align}
being the cut-off distance from the satellite.

\begin{figure*}[t!]
    \begin{equation} \label{result}
       E_h = \frac{2 \mu \, \tan^{-1} \left[ \frac{(H^2 + 2HR + 2R^2 + 2R(H + R)\cos(\phi))\tan\left[\frac{\omega T}{2}\right]}{\sqrt{H^4 + 4h^3R + 6h^2R^2 + 4HR^3 + 2R^4 - 2r^2(H + R)^2\cos[2\phi]}} \right]}{\omega \sqrt{H^4 + 4H^3R + 6H^2R^2 + 4HR^3 + 2R^4 - 2R^2(H + R)^2\cos[2\phi]}}
    \end{equation}
\hrulefill
\end{figure*}

Consequently, As shown in Fig. \ref{fig:sys}, using \eqref{p_r2} and \eqref{distance}, the total amount of energy $E_h$ harvested by the WCD for a charging duration $T$ spanning the charging visibility region during which $P_r\geq P_{th}$, can be defined as
\begin{align} \label{integral}
&E_h =  \int_{t=0}^{T} \modification{P_h(t)} dt \\
&= \mu  \int_{t=0}^{T}\frac{1}{(R + H)^2 + R^2 - 2 (R+H) R \cos \phi \cos(\omega t)} dt,
\end{align}
where \begin{align} \label{T}
T=\theta_c/\omega,
\end{align}
and $\mu=\eta_h P_t G_T G_R \frac{\lambda^2}{(4\pi)^2} \alpha_s \beta_s$ with $\eta_h$ being the WCD harvesting efficiency\footnote{\modification{Note that, the atmospheric losses due to absorption, rain and fog are considerable only for frequencies higher than 10 GHz, which are not typical for satellite WET setups. Moreover, the  Earth limb blockage exists only for very low elevation angles (e.g, $5^\circ$), which is beyond our practical setup due to the existence of the cut-off angle $\theta_c$}}. Solving this integral using (2.553.3) and (2.553.4) in \cite{gradshteyn}, we obtain the closed-form expression on the top of the this page. The details of the solution are depicted in Appendix A. This formula further reduces to the following simpler expression for an azimuth angle $\phi=0$ (i.e, when the satellite orbit and the WCD are in the same plan)
\begin{equation}
    E_h (\phi=0) = \mu \frac{\pi - 2 \tan^{-1} \left( \frac{H \cot \left( \frac{wT}{2} \right)}{H + 2R} \right)}{H^2 \omega + 2HR\omega},
\end{equation}
which acts as an upper-bound.

A special case of energy harvesting is to make use of the merits of beamforming via maximum ratio transmission \cite{MRT} by considering $M$ antennas per satellite. In such case, it is reasonable to assume that if the satellites are close to each other (i.e, satellite grid), then the channel coefficient between them and the ground device, would be almost constant due to the large distance between the satellites and the WCD. According to (10) and (13) in \cite{MRT}, the amount of harvested energy is multiplied by the square of the total number of antennas $(M N^2)$ and the total amount of harvested energy at the earth device is theoretically upper bounded by
\begin{equation}\label{eq:harvestedEnergy}
    E_{h_u} = 2(MN^2) \mu \frac{\pi - 2 \tan^{-1} \left( \frac{H \cot \left( \frac{wT}{2} \right)}{H + 2R} \right)}{H^2 w + 2HRw}.
\end{equation}
\begin{remark}
Interestingly, \eqref{eq:harvestedEnergy} reveals that increasing the number of satellites or the number of antennas per satellite have a similar effect on boosting the upper bound of the harvested energy. Thus, it is possible to reduce the number of satellites in a satellite grid by equipping the satellites with more antennas, which would save the costs of launching larger number of satellites per grid from an economic point of view. However, more studies are required on the practicality of antenna densification of satellites and the realistic performance in that case.
\end{remark}

\subsection{Charging Efficiency}

Finally, defined as the ratio between the actual harvested energy and the available energy from the satellite grid during its visibility path (i.e, when $P_{th}=0$), the charging efficiency $\eta_c=\frac{E_h}{E_h(P_{th}=0)}$ indicates the portion of energy that is actually harvested during the satellites visibility range due to the circuit threshold limitations.

\modification{\subsection{Phase Misalignment}
Taking into consideration the satellite phase shifts according to \cite{Phase_paper}, the phase variation at the WCD could be modeled as $\left[e^{j\psi_1}, \ldots, e^{j\psi_N}\right] $, where $\psi_n$ denotes the phase error at satellite $n$. For i.i.d.\ phase errors with $\psi_n \sim \mathcal{N}(0,\sigma_\psi^2)$, the cross power terms are attenuated by the factor $e^{-\sigma_\psi^2}$, where $\sigma_\psi^2$ is the variance of the phase error distribution.}

\section{Numerical Results}\label{sec:results} 

\begin{table}[t!] \label{tab:paramters}
\centering
\caption{Simulation parameters}
\label{tab:Par}
\begin{tabular}{cc|cc}
\toprule
\textbf{Parameter}                                    & \textbf{Value} & \textbf{Parameter}                                    & \textbf{Value}\\ \midrule
$R$ & $6378$ Km & $H$ & $200$ Km \\
$G$ & $6.67\!\times10^{-11}$$\frac{m^3}{Kg \ s^2}$ & $M_e$ & $5.97\! \times 10^{24}$ Kg \\
$G_T$ & $50 \: \text{dB}$ & $G_r$ & $10 \: \text{dB}$ \\

$\Omega$ & 1.29 & $f$ & $868$ MHz \\
$m$ & 19.4  & $P_t$ & 40~\text{dBm} \\
$b_o$ & 0.158 &   $\eta_h$ & $70\%$ \\
\bottomrule
\end{tabular} 
\end{table}

\modification{We consider a grid of LEO satellites operating within the L-Band (i.e, 858 MHz).} Each satellite is equipped with four antenna elements, moving together while charging a WCD. Moreover, we consider that the charging window is determined by both the sensitivity of the WCD and the horizon point. For each value of $P_{th}$ depicted in the following figures, we calculate $d_{c}$ and $\theta_c$, then $T$ from \eqref{theta_th}, \eqref{d_th}, \eqref{T}, respectively. Note that, $T$ spans the satellite charging path as shown in Fig. \ref{fig:sys}. Table \ref{tab:Par} illustrates all the simulation parameters in all simulations unless stated otherwise. \modification{Note that the simulation parameters including channel parameters are empirically chosen within the practical ranges of LEO satellites operating in light shadowing as in well-established works such as \cite{Anna}. Finally, the parameters of the WCD are selected based on the measured characteristics of the circuit in \cite{PowercastP2110B2024}.}

\begin{figure}[t!]
    \centering
    \includegraphics[width=0.97\linewidth]{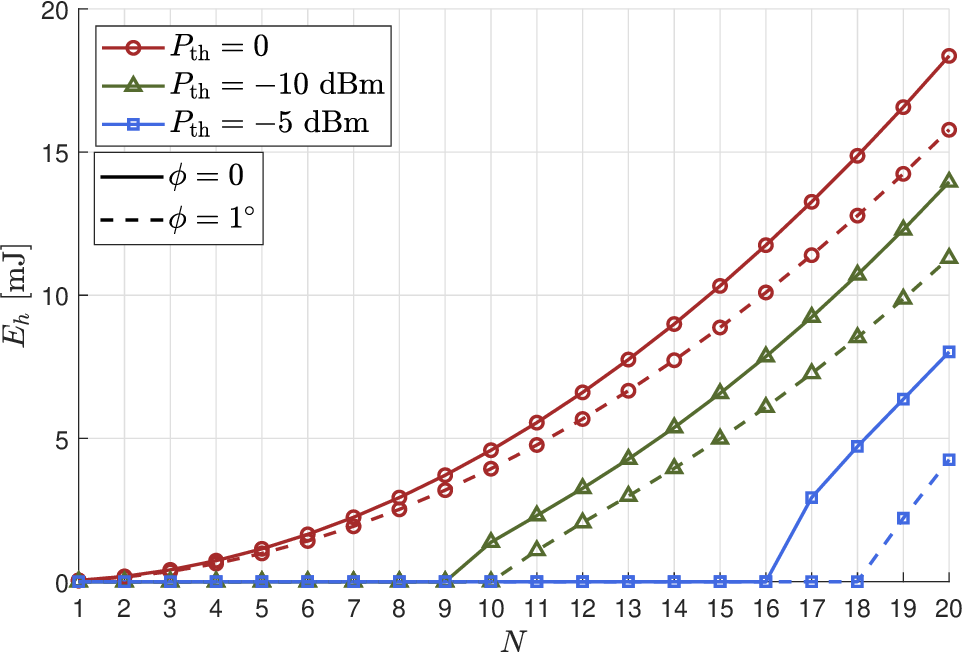}
    \caption{Harvested energy vs the number of satellites for different WCD sensitivities.}
    \label{fig:rEnergyVSNumSat}
\end{figure}

Fig.~\ref{fig:rEnergyVSNumSat} depicts the harvested energy at the WCD as a function of the constellation size, under varying sensitivity thresholds and azimuth angle $\phi$ values. As anticipated from \eqref{eq:harvestedEnergy}, increasing the number of satellites enhances the harvested energy. Apparently, when accounting for practical WCDs with non-ideal sensitivity, a minimum constellation size is required to ensure successful energy harvesting, e.g., $9$ and $16$ for $P_\mathrm{th}=-10~$, $-5~$ dBm respectively, when $\phi = 0^\circ$. This is because increasing the sensitivity of the WCD decreases the duration of the charging window and consequently the achievable harvested energy. \modification{Note that, $P_\mathrm{th} = 0$ means that the circuit is able to sense and harvest energy even if the received power is too small (i.e, $P_r\to 0$). This is considered to be the benchmark ideal case.} Also, observe that when the device is slightly offset (i.e, by only $ 1^\circ$) from satellites' ground track the achievable harvested energy significantly declines. 

\begin{figure}[t!]
    \centering
    \includegraphics[width=0.97\linewidth]{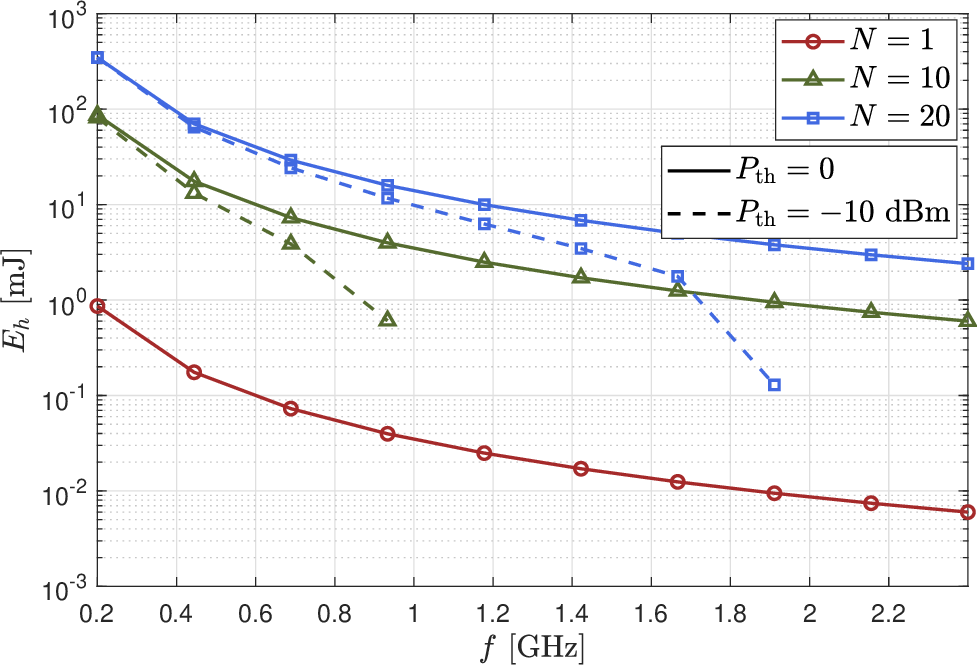}
    \caption{Harvested energy vs the operating frequency for different number of satellites as well as practical and ideal WCD circuits.}
    \label{fig:rEnergyVSFreq}
\end{figure}

Fig.~\ref{fig:rEnergyVSFreq} shows how the harvested energy is affected by the operating frequency due to the higher free-space path losses for higher frequencies. The figure elucidates that this effect is more pronounced for smaller constellation sizes, whereas larger constellations enhance spatial diversity and thus improve the total harvested energy. Moreover, when accounting for the sensitivity of the WCD, the achievable harvested energy is further constrained, rendering certain portions of the frequency spectrum infeasible for energy harvesting depending on the number of satellites being used. For instance, the maximum feasible frequency for 10 satellites is 950 MHz, while it is extended to 1.9 GHz for 20 satellites.

\begin{figure}[t!]
    \centering
    \includegraphics[width=0.97\linewidth]{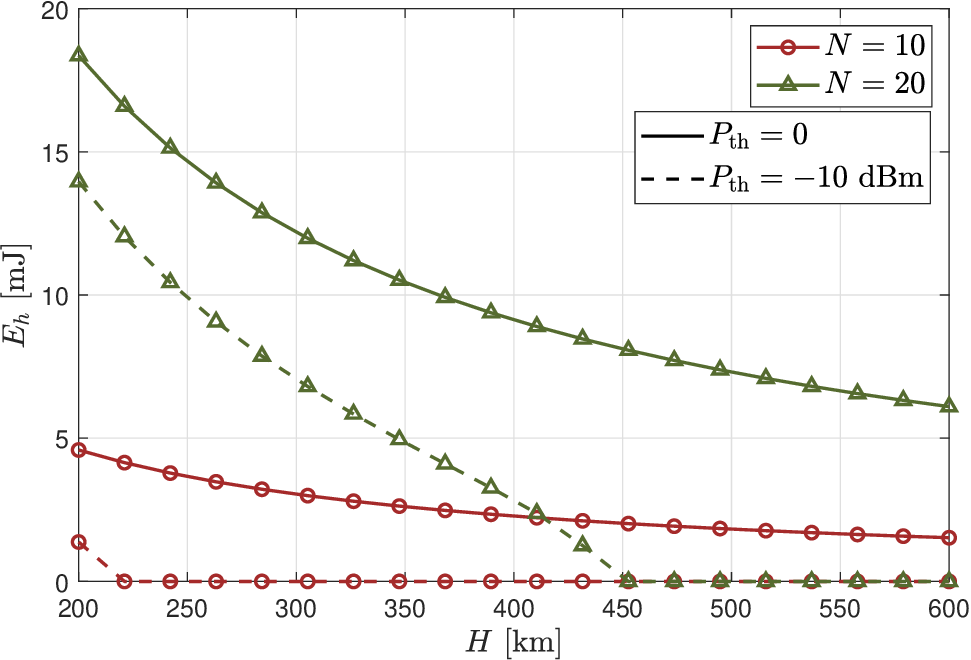}
    \caption{\modification{Harvested energy vs the height of the swarm for different number of satellites and practical and ideal WCD circuits.}}
    \label{fig:visibilityVSHeight}
\end{figure}

Fig.~\ref{fig:visibilityVSHeight} depicts how increasing the number of satellites can enable charging from higher distances. For practical WCDs with non-ideal sensitivity, the maximum allowable (i.e, feasible) constellation height becomes a function of the constellation size. For instance, as observed from the figure, the maximum feasible height for charging a WCD \modification{using} a grid of 10 satellites is \modification{$220~$km}, whereas a grid of $20$ satellites extends the feasible height to nearly \modification{$440~$km}. \modification{In the most favorable scenario, simultaneous charging by $20$ at $d(t) = H = 200~$km yields a received power of approximately $-3~$dBm. This value is well-below the saturation point of the circuit in \cite{PowercastP2110B2024} which is why this operating region has not been considered in \eqref{eq:ehModel}.}

\begin{figure}[htp]
    \centering
    \begin{subfigure}{0.97\columnwidth}
        \centering
        \includegraphics[width=\linewidth]{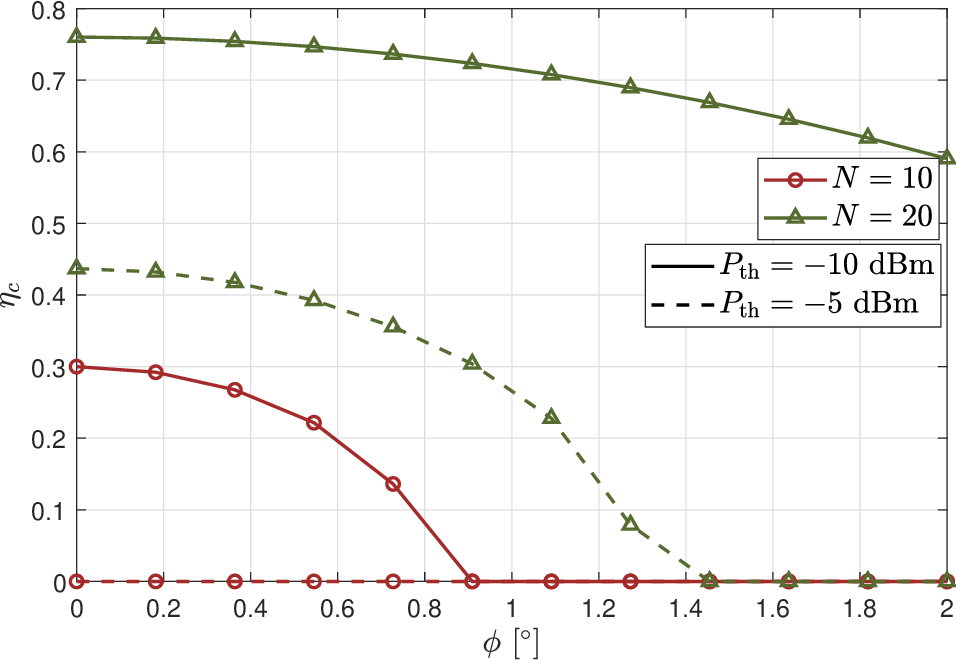}
        \caption{Charging efficiency vs $\phi$ for different number of satellites.}
        \label{fig:efficiencyVSAngle}
    \end{subfigure}
    
    \bigskip 
    
    \begin{subfigure}{0.97\columnwidth}
        \centering
        \includegraphics[width=\linewidth]{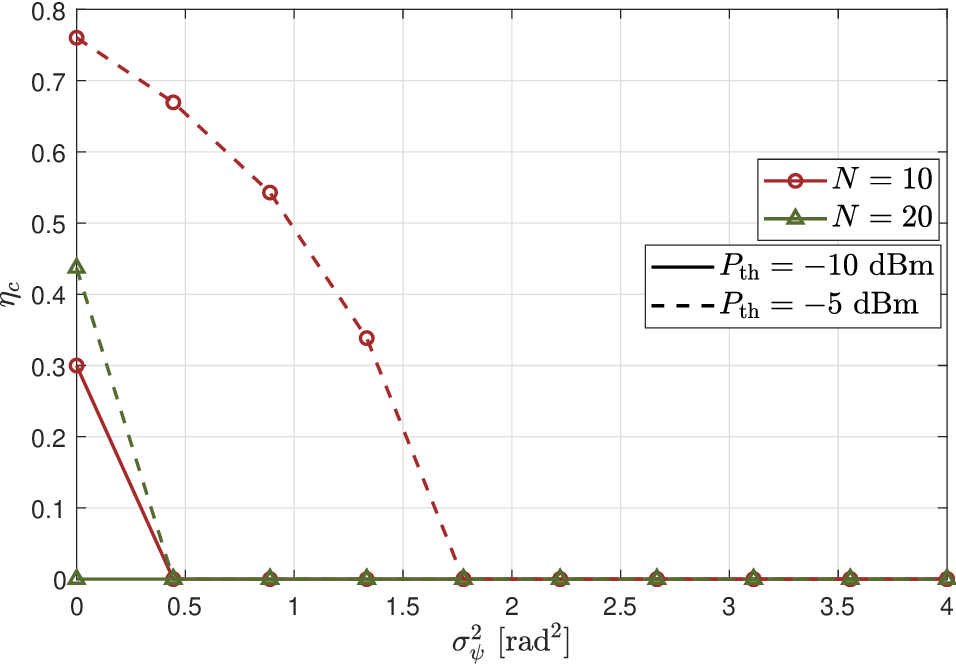}
        \caption{\modification{Charging efficiency vs $\sigma^2_\psi$ for different number of satellites.}}
        \label{fig:phase}
    \end{subfigure}
    
    \caption{Charging efficiency}
    \label{fig:main_caption}
\end{figure}

Fig.~\ref{fig:efficiencyVSAngle} illustrates how the charging efficiency is affected by the satellite azimuth angle $\phi$. \modification{The main takeaway here is that, the charging efficiency is very sensitive to $\phi$ as it decreases rapidly and energy availability is not well exploited when the satellite plane is slightly inclined, specially for higher receiver sensitivity and lower number of satellites.} Therefore, when multiple satellite swarms are available, the swarm whose ground track is closest to the device contributes dominantly to the harvested energy. \modification{Meanwhile, Fig.~\ref{fig:phase} illustrates the effect of phase misalignment on the charging efficiency. We observe that the charging efficiency significantly decays, when the variance of the phase misalignment $\sigma_\psi^2$ becomes higher.}

\section{Conclusions}\label{sec:conclusions} 

This letter presented a framework for space-to-ground energy transmission using satellites, which could be applied to charge IoT or user devices in remote areas or disaster situations. In summary, we suggested a layout, where a group of stacked satellites equipped with multiple antennas are charging a ground WCD, within the visibility range of its orbit. We derived closed form expressions for the amount of energy harvested aided by the maximum ratio transmission gain. The simulation results showed the interplay between the harvested energy, satellite height, number of satellites and transmission frequency for different circuit cut-off setups. Future research directions include but are not limited to exploring laser based charging from space including inter-satellite WET. \modification{Moreover, the model provided in this paper can be extended to joint energy and information transfer from satellites, for instance via modulating the noise statistical properties (i.e, noise modulation).}

\appendices 

\modification{\section{Evaluation of the Integral in \eqref{integral}}
First, define the constants
\begin{align*}
A =& (R+H)^2 + R^2 = H^2 + 2HR + 2R^2, \\
B =& 2R(R+H)\cos\phi .
\end{align*}
Then, the integral becomes
\begin{equation}
E_h=\mu \int_0^T \frac{1}{A - B\cos(\omega t)}dt .
\end{equation}
From trigonometry, we have
\begin{equation}
\cos(\omega t) = \frac{1-\tan^2\left(\frac{\omega t}{2}\right)}{1+\tan^2\left(\frac{\omega t}{2}\right)}.
\end{equation}
Let $u = \tan\left(\frac{\omega t}{2}\right)$. Differentiating, we get
\begin{equation}
dt = \frac{2}{\omega}\frac{du}{1+u^2}.
\end{equation}
Substituting into the integral yields
\begin{align}
E_h=&\frac{2\mu}{\omega}\int_0^T\frac{du}{A(1+u^2) - B(1-u^2)} \\
=&\frac{2\mu}{\omega}\int_0^T\frac{du}{(A-B) + (A+B)u^2}.
\end{align}
According to \cite{gradshteyn}, this integral has a standard form and can be evaluated as
\begin{equation}
\frac{2\mu}{\omega\sqrt{A^2-B^2}}
\tan^{-1}\!\left(
\sqrt{\frac{A+B}{A-B}}u
\right).
\end{equation}
Applying the limits of the integral and expanding the constants yields the expression in \eqref{result}.}


\bibliographystyle{IEEEtran}
\bibliography{references}
\end{document}